\def\Title{Ambiguities in the derivation of retrodictive probability}
 \def\arXiv{quant-ph/0207101}
\documentclass[aps,pra,nobibnotes,nofootinbib,twocolumn,twoside,noshowpacs,fleqn,draft]{revtex4}%
 \usepackage[tbtags,sumlimits]{amsmath}
 \usepackage{amssymb}
 \usepackage{xspace}

 \newcommand{\RefEqn}[1]{Eq.~\eqref{#1}}
 \newcommand{\RefEqns}[1]{Eqs.~\eqref{#1}}

 \newcommand{\RefCite}[1]{Ref.~\onlinecite{#1}}

 \newcommand{\DefEq}{\equiv}%{\doteq}
 \newcommand{\abs}[1]{\ensuremath{\left\vert#1\right\vert}}
 \newcommand{\Sum}[1]{\ensuremath{\sum_{#1}}}
 
 \newcommand{\set}[1]{\ensuremath{{\left\{\,#1\,\right\}}}}%
 \newcommand{\setsuch}[2]{\ensuremath{\big\{\,#1\,\big|\,#2\,\big\}}}%

 \newcommand{\Orj}[2]{\ensuremath{{\textstyle\bigvee}_{\!#1}\,{#2}}}

 \newcommand{\ket}[1]{\ensuremath{\vert\,{#1}\,\rangle}}
 \newcommand{\bra}[1]{\ensuremath{\langle\,#1\,\vert}}
 \newcommand{\braket}[2]{\ensuremath{\langle#1\,\vert\,#2\rangle}}% <#1|#2>
 \newcommand{\proj}[1]{\ensuremath{\ket{{#1}}\bra{{#1}}}}

 \newcommand{\bRho}{\pmb{\rho}}

 \renewcommand{\Pr}[1]{\ensuremath{{\rm Pr}\bigl(\,{#1}\,\bigr)}}
 \newcommand{\Prob}[2]{\ensuremath{{\rm Pr}\bigl(\,{#1}\bigm|#2\,\bigr)}}

 \newcommand{\p}[1]{\ensuremath{p_{#1}}\xspace}
 \newcommand{\q}[1]{\ensuremath{q_{#1}}\xspace}

 \newcommand{\MP}{\ensuremath{M_P}\xspace}

 \newcommand{\MPs}[1]{\ensuremath{M_P^{[#1]}}\xspace}
 \newcommand{\ps}[2]{\ensuremath{p_{#2}^{[#1]}}\xspace}
 \newcommand{\qs}[2]{\ensuremath{q_{#2}^{[#1]}}\xspace}

\begin{document}
 \makeatletter
 \def\ps@titlepage{%
   \renewcommand{\@oddfoot}{}%
   \renewcommand{\@evenfoot}{}%
   \renewcommand{\@oddhead}{\hfill\arXiv}
   \renewcommand{\@evenhead}{}}
 \makeatother

%%%%%%%%%%%%%%%%%%%%%%%%%%%%%%%%%%%%%%%%%%%%%%%%%%%%%%%%%%%%%%%%%%%%%%%%%%%%%%%%%%%%%%%
\title[\Title]
      {\Title}
%%%%%% PERSONAL %%%%%%%%%%%%%%%%%%%%%%%%%%%%%%%%%%%%%%%%%%%%%%%%%%%%%%%%%%%%%%%%%%%%%%%%
\author{K.~A.~Kirkpatrick}
\email[E-mail: ]{kirkpatrick@physics.nmhu.edu}
%\homepage[]{}
\affiliation{New Mexico Highlands University, Las Vegas, New Mexico 87701}
%%%%%%%%%%%%%%%%%%%%%%%%%%%%%%%%%%%%%%%%%%%%%%%%%%%%%%%%%%%%%%%%%%%%%%%%%%%%%%%%%%%%%%%%
%%%%%% ABSTRACT %%%%%%%%%%%%%%%%%%%%%%%%%%%%%%%%%%%%%%%%%%%%%%%%%%%%%%%%%%%%%%%%%%%%%%%%
\begin{abstract}
The derivation of the quantum retrodictive probability formula involves an error, an
ambiguity. The end result is correct because this error appears twice, in such a way as
to cancel itself. In addition, however, the usual expression for the probability itself
contains the same ambiguity; this may lead to errors in its application. A generally
applicable method is given to avoid such ambiguities altogether.
\end{abstract}
% insert suggested PACS numbers in braces on next line
% \pacs{03.65.Bz, 01.70.+w}

%%%%%%%%%%%%%%%%%%%%%%%%%%%%%%%%%%%%%%%%%%%%%%%%%%%%%%%%%%%%%%%%%%%%%%%%%%%%%%%%%%%%%%%%
 \maketitle
%%%%%%%%%%%%%%%%%%%%%%%%%%%%%%%%%%%%%%%%%%%%%%%%%%%%%%%%%%%%%%%%%%%%%%%%%%%%%%%%%%%%%%%%
%% centered short title in each header:
 \makeatletter\markboth{\hfill\@shorttitle\hfill}{\hfill\@shorttitle\hfill}\makeatother
 \pagestyle{myheadings}
%%%%%%%%%%%%%%%%%%%%%%%%%%%%%%%%%%%%%%%%%%%%%%%%%%%%%%%%%%%%%%%%%%%%%%%%%%%%%%%%%%%%%%%%
%%% BODY OF DOCUMENT %%%%%%%%%%%%%%%%%%%%%%%%%%%%%%%%%%%%%%%%%%%%%%%%%%%%%%%%%%%%%%%%%%%
%%%%%%%%%%%%%%%%%%%%%%%%%%%%%%%%%%%%%%%%%%%%%%%%%%%%%%%%%%%%%%%%%%%%%%%%%%%%%%%%%%%%%%%

\section{Introduction} In \RefCite{ABL}, Aharonov, Bergmann, and Lebowitz (ABL)
examined, in quantum mechanics, the following problem:
\begin{quote}
After a system is prepared, the system variable $P$ is observed, followed by the
observation of the variable $Q$. It is desired to determine (``retrodict'')
$\Prob{\p{j}}{\q{k}}$, the probability of the earlier event $P=\p{j}$ given the later
occurrent fact $Q=\q{k}$.
\end{quote}
They obtained, for a system initially prepared in the state $\bRho$, the retrodiction
formula
\begin{equation}\label{E:QResult}
  \Prob{\p{j}}{\q{k}}=\frac{\abs{\braket{\q{k}}{\p{j}}}^2 \bra{\p{j}}\bRho\ket{\p{j}}}%
                      {\Sum{s}\abs{\braket{\q{k}}{\p{s}}}^2 \bra{\p{s}}\bRho\ket{\p{s}}}.
\end{equation}
Their derivation of this (correct) expression contains an error which appears twice,
canceling itself. Further, the probability expression on the left side itself is
ambiguous, which may lead to confusion in application.

\section{The derivation}
The following derivation is implicit in \RefCite{ABL}; it appears somewhat more
explicitly in \RefCite{AharonovV91} and in \RefCite{BarnettPJ00}.

Ordinarily one would solve this problem quite directly, using Bayes's Formula:
\begin{equation}\label{E:First}
 \Prob{\p{j}}{\q{k}}=\frac{\Prob{\q{k}}{\p{j}}\Pr{\p{j}}}{\Pr{\q{k}}}.
\end{equation}
But this results in
\begin{equation}\label{E:QResultWrong}
  \Prob{\p{j}}{\q{k}}=\frac{\abs{\braket{\q{k}}{\p{j}}}^2 \bra{\p{j}}\bRho\ket{\p{j}}}%
                      {\bra{\q{k}}\bRho\ket{\q{k}}},
\end{equation}
which ``doesn't work'' in quantum mechanics! (Consider a spin-1/2 system prepared as
$\bRho=\proj{z+}$, with $\ket{q_k}=\ket{z-}$ and $\ket{p_{1,2}}=\ket{y+,-}$.)

Of course, ABL recognized this, and assumed that the event \q{k} arises following the
events \set{\p{j}}; for the denominator of \RefEqn{E:First}, they used the
marginal-probability formula,
\begin{equation}\label{E:Marginal}
  \Pr{\q{k}}=\Sum{s}\Pr{\p{s}\wedge\q{k}}=\Sum{s}\Prob{\q{k}}{\p{s}}\Pr{\p{s}}.
\end{equation}
In quantum-mechanical terms, this is
\begin{equation}\label{E:MarginalQM}
  \Pr{\q{k}}=\Sum{s}\abs{\braket{\p{s}}{\q{k}}}^2\bra{\p{s}}\bRho\ket{\p{s}},
\end{equation}
which, in \RefEqn{E:First}, results in the ABL formula \RefEqn{E:QResult}. (Further,
\RefEqn{E:Marginal} leads to the classical retrodiction expression
\begin{equation}\label{E:Result}
   \Prob{\p{j}}{\q{k}}=\frac{\Prob{\q{k}}{\p{j}}\Pr{\p{j}}}%
                              {\Sum{s}\Prob{\q{k}}{\p{s}}\Pr{\p{s}}},
\end{equation}
which directly implies the quantum ABL Formula  \RefEqn{E:QResult}.)

However, as Margenau\cite{Margenau63a} noted, there is something wrong with
\RefEqn{E:MarginalQM}: for the pure state $\bRho=\proj{\Psi}$, \RefEqn{E:MarginalQM} is
\begin{equation}
 \abs{\braket{\q{k}}{\Psi}}^2=%
     \Sum{s}\abs{\braket{\q{k}}{\p{s}}}^2\abs{\braket{\p{s}}{\Psi}}^2,
\end{equation}
which, for $\ket{\Psi}\notin\set{\ket{\p{j}}}$, is impossible.

Thus \RefEqn{E:First} and \RefEqn{E:Marginal}, which seem to be correct in ordinary
probability theory, both fail in quantum mechanics. It has been
suggested\cite{Margenau63a} that this implies a special, different ``quantum
probability.'' However, the reality is more pedestrian: Quantum mechanics is a
probability theory of sequences of events in systems of several variables. In such
sequences issues arise which are unfamiliar in, but not foreign to, ordinary probability
theory.\cite{Kirkpatrick:Quantal} In the above derivation these issues have not been
dealt with, with the result that neither \RefEqn{E:First} nor \RefEqn{E:Marginal} is
correct.

In order to see the source of these errors, let us rewrite \RefEqn{E:First}, using a
more-careful notation which denotes the ordinal position of each event by bracketed
superscripts:
\begin{equation}\label{E:FirstOrdinal}
 \Prob{\ps{1}{j}}{\qs{2}{k}}=\frac{\Prob{\qs{2}{k}}{\ps{1}{j}}\Pr{\ps{1}{j}}}%
                                  {\Pr{\qs{2}{k}}}. \tag{{\ref{E:First}}$^{\prime}$}
\end{equation}
The denominator of \RefEqn{E:FirstOrdinal} is obviously ambiguous: what
\textsl{event}$^{[1]}$ precedes \qs{2}{k}? I have shown, in
\RefCite{Kirkpatrick:Quantal}, that this ambiguity is not innocent, even in ordinary
non-quantal probability---the expression \Pr{\qs{2}{k}} may be undefinable (and is
undefinable in quantum mechanics). The same ambiguity appears on the left side of
\RefEqn{E:Marginal}, which is thus also incorrect.

\section{A derivation without ambiguities}
Let us derive an expression for Bayes's Formula which avoids this ambiguity; this will
allow us to show that \RefEqn{E:Result} (and hence \RefEqn{E:QResult}) is correct,
showing that the errors in \RefEqns{E:First} and \eqref{E:Marginal} cancel one-another
in the ABL derivation of \RefEqn{E:Result}.

Given the complete, disjoint set of values \set{\p{j}}, we introduce the notation
$\MP\DefEq\Orj{s}{\p{s}}$; then $\p{j}\equiv\p{j}\wedge\MP$, and
\begin{equation}\label{E:pMq}
 \Pr{\ps{1}{j}\wedge\qs{2}{k}}=\Pr{\ps{1}{j}\wedge\MPs{1}\wedge\qs{2}{k}}.
\end{equation}
The derivation of Bayes's Formula involves applying the definition of conditional
probability to the conjunction \Pr{\ps{1}{j}\wedge\qs{2}{k}} in either order: First,
using \RefEqn{E:pMq},
\begin{align}\label{E:Straight}
  &\Pr{\ps{1}{j}\wedge\qs{2}{k}}\notag\\
  &\qquad=\Prob{\qs{2}{k}}{\ps{1}{j}\wedge\MPs{1}}\Pr{\ps{1}{j}\wedge\MPs{1}},\\
  \intertext{which simplifies to (the expected)}
  &\Pr{\ps{1}{j}\wedge\qs{2}{k}}=\Prob{\qs{2}{k}}{\ps{1}{j}}\Pr{\ps{1}{j}}.
\end{align}
Second, again using \RefEqn{E:pMq},
\begin{align}
 &\Pr{\ps{1}{j}\wedge\qs{2}{k}}\notag\\
  &\qquad=\Prob{\ps{1}{j}}{\MPs{1}\wedge\qs{2}{k}}\Pr{\MPs{1}\wedge\qs{2}{k}};
%\end{align}
\intertext{since $\Pr{\MP}=1$,}
%\begin{align}
 &\Prob{\ps{1}{j}}{\MPs{1}\wedge\qs{2}{k}}=\Prob{\MPs{1}\wedge\ps{1}{j}}{\qs{2}{k}}\notag\\
 &\qquad=\Prob{\ps{1}{j}}{\MPs{1}\wedge\qs{2}{k}},
%\end{align}
\intertext{so}
%\begin{equation}\label{E:Reverse}
  &\Pr{\ps{1}{j}\wedge\qs{2}{k}}=%
  \Prob{\ps{1}{j}}{\MPs{1}\wedge\qs{2}{k}}\Pr{\MPs{1}\wedge\qs{2}{k}}.\label{E:Reverse}
%\end{equation}
\end{align}
Combining \RefEqns{E:Straight} and \eqref{E:Reverse}, we obtain, in place of
\RefEqn{E:First}, the correct expression of Bayes's Formula,
\begin{equation}\label{E:FirstCorrect}
 \Prob{\ps{1}{j}}{\MPs{1}\wedge\qs{2}{k}}=\frac{\Prob{\qs{2}{k}}{\ps{1}{j}}\Pr{\ps{1}{j}}}%
                                  {\Pr{\MPs{1}\wedge\qs{2}{k}}}.
\end{equation}
The no-longer ambiguous denominator expands to
\begin{align}
 &\Pr{\MPs{1}\wedge\qs{2}{k}}=\Sum{s}\Pr{\ps{1}{s}\wedge\qs{2}{k}}\notag\\
 &\qquad=\Sum{s}\Prob{\qs{2}{k}}{\ps{1}{s}}\Pr{\ps{1}{s}};\label{E:MarginalCorrect}
\end{align}
replacing this in \RefEqn{E:FirstCorrect}, we obtain
\begin{equation}\label{E:ResultCorrect}
 \Prob{\ps{1}{j}}{\MPs{1}\wedge\qs{2}{k}}=\frac{\Prob{\qs{2}{k}}{\ps{1}{j}}\Pr{\ps{1}{j}}}%
                                  {\Pr{\MPs{1}\wedge\qs{2}{k}}},
\end{equation}
thus \RefEqn{E:Result} (and hence,  \RefEqn{E:QResult}) is correct except for the
ambiguity on the left.

\section{Comments}%

We see from \RefEqn{E:FirstCorrect} that the error in  \RefEqn{E:FirstOrdinal} is indeed
the ambiguity in the denominator \Pr{\qs{2}{k}}: the missing \textsl{event}$^{[1]}$ is
\MP, the ignored observation of $P$: the denominator may be written simply
\Prob{\qs{2}{k}}{\MPs{1}}, ``the probability of \q{k} following the ignored complete
observation of $P$.''

Further, the left side of \RefEqn{E:QResult} should be written
\Prob{\ps{1}{j}}{\MPs{1}\wedge\qs{2}{k}}: the ignored complete observation of $P$ must
be explicitly accounted for. Why is this important? First, rotate \set{\ket{\p{s}}} about
\ket{\p{1}} to get \setsuch{\ket{{p_j}'}}{\ket{{p_1}'}=\ket{\p{1}}}, the eigenstates of a
variable $P'$ which has its $j=1$-vector in common with $P$; ignoring \emph{this}
complete observation yields a different value:
$\Prob{{\p{1}}'}{\q{k}}\neq\Prob{\p{1}}{\q{k}}$. (This is the ``something very curious''
which arises in \RefCite{AlbertAD85}.) Second, the observation needn't be complete: for
example, merely observe ``\p{j} or not \p{j}''; in this case the denominator of
\RefEqn{E:QResult} becomes
\[\abs{\braket{\q{k}}{\p{j}}}^2\bra{\p{j}}\bRho\ket{\p{j}}+
 \Sum{ss'\neq j}\braket{\p{s'}}{\q{k}}\braket{\q{k}}{\p{s}}\bra{\p{s}}\bRho\ket{\p{s'}}\]
This leads to the ``Three-Box Paradox''\cite{AharonovV91}, which is surprising partly
because of a failure to explicitly note the difference between the condition
$(\p{1}\vee\p{2}\vee\p{3})^{[1]}\wedge\qs{2}{k}$ and the condition
$(\p{1}\vee\neg\p{1})^{[1]}\wedge\qs{2}{k}$.
%This ambiguity lead to certain results claimed in \RefCite{AlbertAD85}, results
%seemingly abandoned by \RefCite{AlbertAD86}, and which lead rather directly to the
%Three-Box example of \RefCite{AharonovV91}.

The derivation presented by Aharonov, Bergmann, and Lebowitz\cite{ABL} is implicit: they
start with \RefEqn{E:First} (which appears as the first part of their Eq.~(2.4)). The
result \RefEqn{E:QResult} is equivalent to the quantum expressions in their Eqs.~(2.4)
and (2.5). The transition from \RefEqn{E:First} to \RefEqn{E:QResult} is done in a single
step, without comment, using the quantum equivalent of \RefEqn{E:Marginal}. Thus the
ambiguous \Pr{\qs{2}{k}} appears at the beginning of their derivation, and then simply
vanishes. The cancelation of these errors is more good fortune (and good intuition) than
good physics; Margenau was not so lucky.

\RefEqn{E:MarginalCorrect} is the marginal probability identity appropriate to summing
over the \emph{earlier} event. It is interesting that, deriving a form of the
marginal-probability formula, Ballentine \cite{Ballentine86} used a technique very
similar to the above, but the \MP, after being introduced, was dropped, resulting in the
incorrect \RefEqn{E:Marginal}.

Quantum mechanics involves the classical probability of sequences of events involving
more than one variable. Few treatments of probability deal with such sequences; the
resulting unfamiliarity has lead to numerous errors in the understanding of the
quantum-mechanical probability formulas. For the purpose of extending our fundamental
understanding, the various formal approaches to measurement (involving POMs, POVMs,
effects and operations) are inadequate: based entirely on the the Hilbert-space
formalism, their connection with probability theory is loose and ill-understood.

%%%%%%%%%%%%%%%%%%%%%%%%%%%%%%%%%%%%%%%%%%%%%%%%%%%%%%%%%%%%%%%%%%%%%%%%%%%%%%%%%%%%%%%%
%%% END OF BODY %%%%%%%%%%%%%%%%%%%%%%%%%%%%%%%%%%%%%%%%%%%%%%%%%%%%%%%%%%%%%%%%%%%%%%%%
%%%%%%%%%%%%%%%%%%%%%%%%%%%%%%%%%%%%%%%%%%%%%%%%%%%%%%%%%%%%%%%%%%%%%%%%%%%%%%%%%%%%%%%%

%\begin{acknowledgments}
% put your acknowledgments here.
%\end{acknowledgments}

%%%%%%%%%%%%%%%%%%%%%%%%%%%%%%%%%%%%%%%%%%%%%%%%%%%%%%%%%%%%%%%%%%%%%%%%%%%%%%%%%%%%%%%%
%% BIBLIOGRAPHY %%%%%%%%%%%%%%%%%%%%%%%%%%%%%%%%%%%%%%%%%%%%%%%%%%%%%%%%%%%%%%%%%%%%%%%%
%%%%%%%%%%%%%%%%%%%%%%%%%%%%%%%%%%%%%%%%%%%%%%%%%%%%%%%%%%%%%%%%%%%%%%%%%%%%%%%%%%%%%%%
%\vfill
 \renewcommand{\refname}{\sc References}%xxx
 \footnotesize%

%%%%%%%%%%%%%%%%%%%%%%%%%%%%%%%%%%%%%%%%%%%%%%%%%%%%%%%%%%%%%%%%%%%%%%%%%%%%%%%%%%%%%%%

%% \bibliographystyle{myapsrev}
% \bibliographystyle{ajp}
% \bibliography{JAbbrevs,QM\NotesBib}

%% For submission, comment out previous lines, follow instructions at "Footnotes in
%% Bibliography," above, and copy .bbl here:

%%%%%%%%%%%%%%%%%%%%%%%%%%%%%%%%%%%%%%%%%%%%%%%%%%%%%%%%%%%%%%%%%%%%%%%%%%%%%%%%%%%%%%%%
%%% END DOCUMENT %%%%%%%%%%%%%%%%%%%%%%%%%%%%%%%%%%%%%%%%%%%%%%%%%%%%%%%%%%%%%%%%%%%%%%%
%%%%%%%%%%%%%%%%%%%%%%%%%%%%%%%%%%%%%%%%%%%%%%%%%%%%%%%%%%%%%%%%%%%%%%%%%%%%%%%%%%%%%%%%
\end{document}